\documentclass[conference]{IEEEtran}
\usepackage{subfigure}
\usepackage{cite}
\usepackage{amsmath}
\usepackage{cite}
\usepackage{graphicx}
\usepackage{epstopdf}
\usepackage{amsfonts,amsmath,amssymb}
\usepackage{graphicx}
\usepackage{url}
\usepackage{bm}
\usepackage{bbm}
\usepackage{subfigure}
\usepackage{stfloats}
\usepackage{subeqnarray}
\usepackage{verbatim}
\usepackage{multirow}
\usepackage{makecell}

\usepackage{cite,graphicx,amsmath,amssymb,cite,algorithm}

\begin{document}
\newtheorem{theorem}{\textbf{Theorem}}
\newtheorem{proposition}{\textbf{Proposition}}
\newtheorem{corollary}{\textbf{Corollary}}
\newtheorem{remark}{Remark}

\title{Establishing Secrecy Region for Directional Modulation Scheme with Random Frequency Diverse Array}
\author{
\IEEEauthorblockN{Shengping~Lv$^{\ast}$, Jinsong~Hu$^{\ast}$, Youjia Chen$^{\ast}$, Zhimeng Xu$^{\ast}$, and Zhizhang (David) Chen$^{\dag}$}
\IEEEauthorblockA{$^{\ast}$College of Physics and Information Engineering, Fuzhou University, Fuzhou, Fujian, 350116, China}
\IEEEauthorblockA{$^{\dag}$Department of Electrical and Computer Engineering, Dalhousie University, Halifax, NS B3H 4R2, Canada}
\IEEEauthorblockA{E-mails:~lvshengping\_fzu@163.com,~\{jinsong.hu, youjia.chen, zhmxu\}@fzu.edu.cn,~z.chen@dal.ca}
}

\maketitle
\begin{abstract}
Random frequency diverse array (RFDA) based directional modulation (DM) was proposed as a promising technology in secure communications to achieve a precise transmission of confidential messages, and artificial noise (AN) was considered as an important helper in RFDA-DM. Compared with previous works that only focus on the spot of the desired receiver, in this work, we investigate a secrecy region around the desired receiver, that is, a specific range and angle resolution around the desired receiver. Firstly, the minimum number of antennas and the bandwidth needed to achieve a secrecy region are derived. Moreover, based on the lower bound of the secrecy capacity in RFDA-DM-AN scheme, we investigate the performance impact of AN on the secrecy capacity. From this work, we conclude that: 1) AN is not always beneficial to the secure transmission. Specifically, when the number of antennas is sufficiently large and the transmit power is smaller than a specified value, AN will reduce secrecy capacity due to the consumption of limited transmit power. 2) Increasing bandwidth will enlarge the set for randomly allocating frequencies and thus lead to a higher secrecy capacity. 3) The minimum number of antennas increases as the predefined secrecy transmission rate increases.
\end{abstract}

%
%
\section{Introduction}
As the fifth-generation (5G) wireless networks coming to reality, people and organizations become more dependent on wireless devices to share secure and private information (e.g., location, wireless payment, e-health). Compared with wired communication, wireless communication facilitates great convenience for people's life in accompany with some secure issues that should not be ignored. The broadcasting nature of wireless communication system may lead to the leakage of confidential information, which is undesirable to preserve security and allows any unauthorized receiver to eavesdrop on the wireless communications. As a complement to high-layer encryption techniques, physical layer security has been widely recognized as a promising way to enhance wireless security by exploiting the characteristics of wireless channels \cite{yangnan2016Safeguarding,zhaonan2016IA,hebiao2016}.

In recent years, there has been a growing research interest on directional modulation (DM), which is an approach launched on multiple-antennas enabled physical layer security technology. It remains the original constellation of the confidential messages at the desired direction while distort the constellation of that at the other directions \cite{Daly2009PA,Jinsong2016robust,TaoHong2018Synthesis,Ryan2018Iterative,WenQin2018Hybrid}. In the existing methods of the literatures, most of the works employed the phased array (PA) to achieve DM scheme.
The aforementioned works studied DM under the assumption that the potential eavesdropper locates at different direction from desired user. However, the potential eavesdropper may be passive and never transmit messages, whose location is not available at the transmitter.
Against this background, frequency diverse array (FDA) aided DM scheme was proposed to realize the secure transmission in the more practical scenarios where the legitimate user and eavesdropper may locate in the same direction but with different ranges from the transmitter. To maximize the secrecy capacity, various of algorithms were proposed \cite{Jinsong2017access,shu2018secure,TongShen2019Two,QianCheng2019WFRFT}.  The authors in \cite{Jinsong2017access} proposed random frequency diverse array (RFDA) based DM scheme with the aid of artificial noise (AN) to achieve two dimensions (i.e., angle and range) secure transmission, based on which the orthogonal frequency division multiplexing (OFDM) and the random subcarrier selection were adopted to further enhance the performance of the secure transmissions~\cite{shu2018secure,TongShen2019Two}. The weighted fractional Fourier transform (WFRFT) with FDA technology was employed to achieve power-efficient multi-beam DM transmissions~\cite{QianCheng2019WFRFT}.

However, these works only focus on the secrecy transmission to the spot of the desired receiver. Generally, the eavesdropper may exist anywhere, a secrecy region is more reasonable and useful than just considering one spot. That is, an area around the desired receiver should be guaranteed safety. For convenient, we named the DM-aided scheme with AN as RFDA-DM-AN and one without AN as RFDA-DM in this work. In this work, our main contributions are summarized below.
\begin{itemize}
\item We first provide beampattern with ellipse function at the desired user. Then, we derived the closed-form expressions of lower bounds on the secrecy capacities of the RFDA-DM-AN scheme and RFDA-DM scheme.

\item The requirements of system resource (i.e., the minimum number of antennas and the bandwidth for random frequency mapping) to establish a secrecy region is investigated.

\item Through numerical studies we show the RFDA-DM scheme outperforms the RFDA-DM-AN scheme when the number of antennas is sufficiently large and transmit power is smaller than a specified value. Our conducted analysis enabled transmitter to switch between these two schemes to achieve the maximum secrecy capacity.
\end{itemize}

The rest of this work is organized as follows. Section II presents a system model and constructs the RFDA aided DM scheme. In section III, the expression of secrecy region around the desired user is derived and  the influence of the minimum number of antennas at transmitter and the bandwidth allocated for random frequency mapping on the safety performance of the proposed scheme are examined. Finally, the numerical illustrations are given in Section IV, and conclusions are drawn in the last section.

\emph{Notation:} Scalars are denoted by italic letters, vectors and matrices are denoted by bold-face lower-case and uppercase letters, respectively.
Given a complex vector or matrix, $(\cdot)^T$, $(\cdot)^H$, $\mathrm{tr}(\cdot)$, and $\|\cdot\|$ denote the transpose, conjugate transpose, trace, and norm, respectively. ${{\bf{I}}_M}$ represents identity matrix with size ${{M{\times}M}}$ and the ${{M{\times}M}}$ all-one matrix is referred to as ${{\bf{E}}_M}$. 

\section{System Model}
\subsection{Considered Scenario and Adopted Assumptions}
\begin{figure}[!t]
    \begin{center}
        \includegraphics[width=0.9\columnwidth]{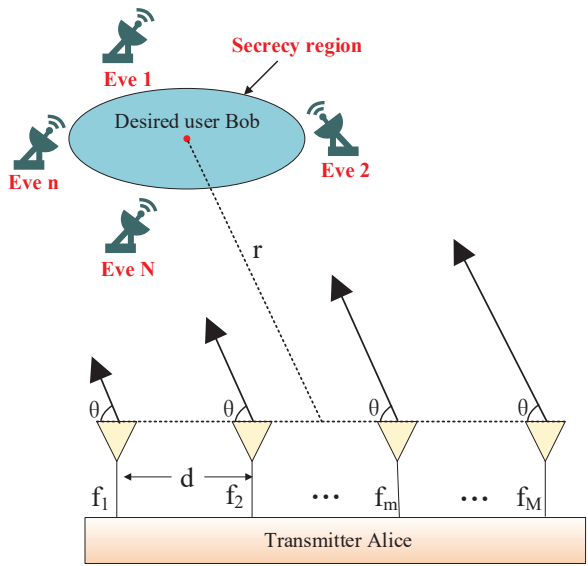}
        \caption{A random frequency diverse array assisted secure communication system}\label{fig1}
    \end{center}
\end{figure}

As shown in Fig.~\ref{fig1}, this work considers a DM-based secure communications network that transmitter Alice wants to send the confidential messages to the desired user Bob, while the potential eavesdropper Eve may exist around Bob and intends to intercept the confidential messages. The transmitter Alice is assumed to adopt the RFDA equipped with an $M$-element antenna array, while all other nodes have a single antenna. The transmitter is assumed to be located at the origin, and the location of receiver denoted by ($r$, ${\theta}$). Moreover, $d$ denotes the element spacing at the transmitter.

Different from the traditional phased array system, the radiation frequency of the $m$-th element of RFDA is given by
\begin{align}
{f_m} = {f_0} + {k_m}\Delta f, ~~~~ m = 1,2,...,M
\end{align}
where ${f_0}$ is the central carrier frequency, $\Delta f$ is the frequency increment, and ${k_m}$ is an independent identically distributed (i.i.d.) random variables. Different distributions of ${k_m}$ lead to different frequency mapping rules used to allocate the carrier frequencies of different elements.
In this work, we define a vector ${\bf{k}}$ to discuss this problem more succinctly and generally as follows:
\begin{align}
{\bf{k}} = {\left[ {{k_1}, {k_2}, ..., {k_M}} \right]^T}.
\end{align}
We defined $K = {\bf{k}}^T{\bf{k}}$, of which the value has correlation to the bandwidth allocated for frequency mapping \cite{Liu2}. As such, the phase shift of $m$-th element is given by \cite{Liu2,Jinsong2017access}
\begin{align}\label{phase shift}
{\Omega _m}(r,\theta ) \approx  - 2\pi \left( {(m - 1)\frac{{{f_0}d\cos \theta }}{c} + {k_m}\Delta f\frac{r}{c}} \right),
\end{align}
where $c$ is the speed of light.

Following \eqref{phase shift}, we can obtain the normalized steering vector of RFDA for a specific location $(r,\theta )$, which is given by
\begin{align}
\mathbf{h}(r,\theta) = \frac{1}{{\sqrt M }}{\left[{e^{j{\Omega _1}(r,\theta )}},{e^{j{\Omega _2}(r,\theta )}},...,{e^{j{\Omega _M}(r,\theta )}}\right]^T}.
\end{align}
In the following, the locations of Bob and the Eve are denoted by $({r_B},{\theta_B})$ and $({r_E},{\theta_E})$, respectively.

\subsection{Direction Modulation with AN}
In this subsection, we first detail the transmissions from Alice to Bob and from Alice to Eve. Then, the associated signal-to-noise ratio (SNR) or signal-to-interference-plus-noise ratio (SINR) is presented.

In DM system, to provide secure communication, AN is generated at the transmitter to distort the signal received by potential eavesdroppers. And hence, the transmitted signal can be expressed as
\begin{align} \label{x_t}
\mathbf{x}_t = \sqrt {\delta {P_t}} \mathbf{v}s + \sqrt {(1 - \delta ){P_t}} \mathbf{w},
\end{align}
where $s$ is the symbol of transmitting confidential message that chosen from the complex signal constellation with average power $\mathbb{E}[|s{|^2}] = 1$, ${P_t}$ is the transmit power of RFDA antennas, $\delta $ is power allocation factor between the confidential signal and AN, $\mathbf{v}$ is the steering vector for the confidential signal, $\mathbf{w}$ is the artificial noise vector. According to \cite{Jinsong2017access}, the expressions of $\mathbf{v}$ and $\mathbf{w}$ are respectively given by
\begin{align}
\mathbf{v}&= \mathbf{h}({r_B},{\theta _B}),
\end{align}
\begin{align}
\mathbf{w}&= \frac{{\left( {{{\bf{I}}_M} - {\bf{h}}({r_B},{\theta _B}){{\bf{h}}^H}({r_B},{\theta _B})} \right){\bf{z}}}}{{\left\| {\left( {{{\bf{I}}_M} - {\bf{h}}({r_B},{\theta _B}){{\bf{h}}^H}({r_B},{\theta _B})} \right){\bf{z}}} \right\|}},
\end{align}
where ${\bf{h}}({r_B},{\theta _B})$ is the steering vector of the transmitter to desired user, ${\bf{z}}$ consists of $M$ i.i.d. circularly-symmetric complex Gaussian random variables with zero-mean and unit-variance, and ${\bf{w}}$ is in the null space vector of $\mathbf{h}^H({r_B},{\theta _B})$. We have $\mathbf{h}^H({r_B},{\theta _B}){\bf{w}}=0$.

According to \eqref{x_t}, the received signal at the desired receiver Bob is given by
\begin{align} \label{y_rB}
y({r_B},{\theta _B}) &= {{\bf{h}}^H}({r_B},{\theta _B}){{\bf{x}}_t} + {n_B}
\notag \\
&= \sqrt {\delta {P_t}} {{\bf{h}}^H}({r_B},{\theta _B}){\bf{h}}({r_B},{\theta _B})s  \notag \\
&~~~+ \sqrt {(1 - \delta ){P_t}} {{\bf{h}}^H}({r_B},{\theta _B}){\bf{w}} + {n_B},
\notag \\
&= \sqrt {\delta {P_t}} s + {n_B}.
\end{align}
where ${n_B}$ denotes the additive white Gaussian noise (AWGN) at Bob ${n_B} \sim {\cal C}{\cal N}\left( {0,\sigma _B^2} \right)$. Hence, the SNR at Bob is given by
\begin{align}
\left( {\mathrm{SNR}} \right)_B=\frac{\delta {P_t}}{\sigma _B^2}  = \delta \mu,
\end{align}
where $\mu  =P_t/\sigma _B^2 $.

The received signal at the eavesdropper Eve is given by
\begin{align} \label{y_rE}
y({r_E},{\theta _E})& = {{\bf{h}}^H}({r_E},{\theta _E}){{\bf{x}}_t} + {n_E} \notag \\
&= \sqrt {\delta {P_t}} {{\bf{h}}^H}({r_E},{\theta _E}){\bf{h}}({r_B},{\theta _B})s  \notag \\
&~~~+ \sqrt {(1 - \delta ){P_t}} {{\bf{h}}^H}({r_E},{\theta _E}){\bf{w}} + {n_E},
\end{align}
where ${n_E} \sim {\cal C}{\cal N}\left( {0,\sigma _E^2} \right)$ is the AWGN at Eve.

Following \eqref{y_rE}, the SINR at Eve is given by
\begin{align}
{\left( {\mathrm{SINR}} \right)}_E& = \frac{{\delta {P_t}{{\left| {{{\bf{h}}^H}({r_E},{\theta _E}){\bf{h}}({r_B},{\theta _B})} \right|}^2}}}{{(1 - \delta ){P_t}{{\left| {{{\bf{h}}^H}({r_E},{\theta _E}){\bf{w}}} \right|}^2} + \sigma _E^2}}  \notag \\
&= \frac{{\delta \mu {{\left| {{{\bf{h}}^H}({r_E},{\theta _E}){\bf{h}}({r_B},{\theta _B})} \right|}^2}}}{{(1 - \delta )\mu {{\left| {{{\bf{h}}^H}({r_E},{\theta _E}){\bf{w}}} \right|}^2} + \varepsilon }},
\end{align}
where $\varepsilon  = {\sigma _E^2}/{\sigma _B^2}$.

\section{Analysis on Secrecy Performance}
In this section, we consider the case where Bob has a secrecy region in the angle and range dimensions. It is assumed that Bob can discover the existence of the Eve immediately once the potential eavesdropper appears in this area. To this end, we introduce mathematical analysis to this system and examine the factors that affect the system's security performance, based on which we can improve the performance. The performance limits of such system is characterized in terms of the secrecy capacity, which is given by
\begin{align}
C&=\max\left\{C_B-C_E, 0\right\},
\end{align}
where $C_B$ and $C_E$ are Bob's channel capacity and Eve's channel capacity, respectively.
\subsection{Mathematical Analysis}
We assume that Bob's offset of the range and angle for secrecy region is $\Delta r$ and $\Delta \theta$, respectively, i.e., the secrecy region around Bob is set $\tilde{r}_B\in [r_B-\Delta r, r_B+\Delta r]\cap \tilde{\theta} _B\in [\theta _B-\Delta \theta, \theta _B+\Delta \theta]$. In this work, we consider the
worst-case scenario when Eve locates at the boundary of the secrecy region. As such, maximum value of the correlation coefficients between the four boundary positions and desired user Bob is given by
\begin{align}
\beta  = \max \left\{ {{{\left| {{{\bf{h}}^H}({r_B} \pm \Delta r,{\theta _B} \pm \Delta \theta ){\bf{h}}({r_B},{\theta _B})} \right|}^2}} \right\},
\end{align}
thus means that the maximum power of confidential signal that can be received by Eve is $\beta$ when Eve is located on the boundary of the secrecy region.
Due to the random location of Eve (outside the secrecy region), we have
\begin{align} \label{hrE_hrB}
\left| {{{\bf{h}}^H}({r_E},{\theta _E}){\bf{h}}({r_B},{\theta _B})} \right|^2= {\left| {\frac{1}{M}\sum\limits_{m = 1}^M e^{jz_m} } \right|^2},
\end{align}
where
\begin{align} \label{zm}
{z_m} &= p{k_m} + q(m-1),
\end{align}
\begin{align} \label{p1}
p &= 2\pi \frac{\Delta f(r_E-r_B)}{c},
\end{align}
\begin{align}  \label{q1}
q &= 2\pi \frac{f_0 d(\cos \theta _E-\cos \theta _B)}{c}.
\end{align}

It should be noted that AN will not affect the power of confidential signal received by Bob and Eve for a given power allocation factor $\delta$. Therefore, in order to achieve the secure transmission, the power of confidential signal outside the secrecy region should be sufficiently small.
i.e., the confidential signal can be transmitted from Alice to Bob safely only when $\left| {{{\bf{h}}^H}({r_E},{\theta _E}){\bf{h}}({r_B},{\theta _B})} \right|^2  \le \beta$. We denote this necessary condition as $\mathbb{S}$.

\begin{theorem}\label{theorem1}
For the condition $\mathbb{S}$, the minimum value of $M$ and $K$ for having secure communication are given by
\begin{align} \label{M_eq}
M_{\min} &= \frac{{{{35.9}^ \circ }c \sqrt {1 - \beta } }}{{\Delta \theta df_0\sin {\theta _B}}},
\end{align}
\begin{align} \label{K_eq}
K_{\min} &= {\left( {\frac{c}{{2\pi \Delta f\Delta r}}} \right)^2}(1 - \beta )M_{\min}.
\end{align}
\end{theorem}
\begin{IEEEproof}
Following \eqref{hrE_hrB} and the condition $\mathbb{S}$, the transmitting beampattern at Eve with the frequency increment vector $\mathbf{k}$ is given by
\begin{align}\label{beampattern}
G({r_E},{\theta _E},{\bf{k}}) = {\left| {\sum\limits_{m = 1}^M {{e^{j{z_m}}}} } \right|^2} \le \beta {M^2}.
\end{align}

Using the second-order Taylor approximation, \eqref{beampattern} can be rewritten as
\begin{align} \label{beampattern_taylor}
G({r_E},{\theta _E},{\bf{k}}) = \sum\limits_{m = 1}^M {\sum\limits_{n = 1}^M {\left[ {1 - \frac{1}{2}{{\left( {{z_m} - {z_n}} \right)}^2}} \right]} }  \le \beta {M^2}.
\end{align}
Substituting ${z_m}$ given in \eqref{zm} into \eqref{beampattern_taylor}, we have
\begin{align}\label{beampattern_taylor_more}
\sum\limits_{m = 1}^M {\sum\limits_{n = 1}^M {{{\left( {{z_m} - {z_n}} \right)}^2}} }  &= \rho _1(\mathbf{k})p^2+2 \rho_2(\mathbf{k}) pq + C q^2 \notag \\
&\ge 2(1 - \beta ){M^2},
\end{align}
where
\begin{align} \label{second}
\rho _1({\bf{k}})&= \left( {\sum\limits_{m = 1}^M {\sum\limits_{n = 1}^M {{{\left( {{k_m} - {k_n}} \right)}^2}} } } \right),
\end{align}
\begin{align} \label{third}
{\rho _2}({\bf{k}}) &= \left( {\sum\limits_{m = 1}^M {\sum\limits_{n = 1}^M {\left( {{k_m} - {k_n}} \right)\left( {m - n} \right)} } } \right),
\end{align}
\begin{align} \label{fourth}
C &= \sum\limits_{m = 1}^M \sum\limits_{n = 1}^M  \left( {m - n} \right) \notag \\
&= \frac{M^2(M^2 - 1)}{6}.
\end{align}

As discussed in \cite{YeziMa2019FDA}, the value of vector $\mathbf{k}$ that translates \eqref{beampattern_taylor_more} into the general form of ellipse equivalently should be the linear combination of eigenvectors corresponding to the maximum eigenvalue of a matrix ${\bf{A}}$, where ${\bf{A}}$ is given by
\begin{align}\label{first}
\bf{A} &= \frac{1}{3}{M^3}\left( {{M^2} - 1} \right) {\mathbf{I}_M} - \frac{2}{3}{M^2}\left( {2{M^2} + 3M + 1} \right) \mathbf{E}_M \notag \\
 &~~~- 4{M^2}{\mathbf{G}} + 2{M^2}(M + 1){\mathbf{P}}.
\end{align}
where $\mathbf{G}$ and $\mathbf{P}$ represent a matrix of size ${{M{\times}M}}$ with value of each element corresponding to the product and the sum of row index $m$ and column index $n$, respectively.

Then $\rho _1(\bf{k})$ and $\rho _2(\bf{k})$ can be further simplified as
\begin{align}\label{rho1ke}
\rho_1(\mathbf{k}) &=  2MK,
\end{align}
\begin{align} \label{rho2ke}
\rho_2(\mathbf{k}) &= 0.
\end{align}

Using \eqref{rho1ke} and \eqref{rho2ke}, we obtain the ellipse function as follows
\begin{align} \label{ellipse_function}
\frac{{{{({r_E} - {r_B})}^2}}}{{{{\left( \Delta {r}^{\dag} \right)}^2}}} + \frac{{({\theta _E} - {\theta _B})}^2}{\left(\Delta {\theta}^{\dag}\right)^2} =  1,
\end{align}
where $\Delta {r}^{\dag}$ and $\Delta {\theta}^{\dag}$ are the solutions of \eqref{ellipse_function}, which are respectively given by \cite{YeziMa2019FDA}
\begin{align} \label{delta_r_constraint}
\Delta {r}^{\dag}   &= \frac{c\sqrt { M (1 - \beta)/K}} {2\pi \Delta f},
\end{align}
\begin{align} \label{delta_theta_constraint}
\Delta {\theta}^{\dag} &= \frac{{35.9}^\circ c  \sqrt{1-\beta}}{Mdf_0\sin {\theta _B}}.
\end{align}
Considering the potential eavesdropper is outside the secrecy region, we have $\Delta {r}\geq\Delta {r}^{\dag}$ and $\Delta {\theta} \geq \Delta {\theta}^{\dag}$. After some algebraic manipulations, results in Theorem~\ref{theorem1} is obtained.
\end{IEEEproof}

\subsection{Secrecy Performance}
In the section II-B, we have discussed the SNR of Bob and SINR of Eve in AN-aided DM scheme, based on which we examine the security performance in this subsection.

First of all, we can get the channel capacity of Bob which is given by
\begin{align} \label{C_B}
{C_B} &= {\log _2}\left( {1 + {{\left( {\mathrm{SNR}} \right)}_B}} \right) \notag \\
&=\log_2(1+\delta \mu).
\end{align}
Similarly, the channel capacity of Eve is given by
\begin{align}
{C_{E - \mathrm{AN}}} &= {\log _2}\left( {1 + {{\left( {\mathrm{SINR}} \right)}_E}} \right) \notag \\
&=\log_2\left(1+\frac{{\delta \mu {{\left| {{{\bf{h}}^H}({r_E},{\theta _E}){\bf{h}}({r_B},{\theta _B})} \right|}^2}}}{{(1 - \delta )\mu {{\left| {{{\bf{h}}^H}({r_E},{\theta _E}){\bf{w}}} \right|}^2} + \varepsilon }}\right)  \notag \\
&\overset{(b)}{\le} {\log _2}\left( 1 + \frac{{\delta \mu \beta }}{{(1 - \delta )\mu \eta(1-\beta)  + \varepsilon }} \right),
\label{C_E_AN}
\end{align}
where (b) is achieved by using the condition $\mathbb{S}$, and $\eta$ is defined as
\begin{align}
\eta=\frac{1}{\mathrm{tr}\left\{\left[\mathbf{I}_M-{\mathbf{h}(r_B, \theta_B)}{\mathbf{h}^H(r_B, \theta_B)}\right]^2\right\}}.
\end{align}
Note that $\eta$ can be easily obtained by the location of Bob (i.e., $(r_B, \theta_B)$), which is assumed to be available at the transmitter.

For the RFDA-DM-AN scheme, the secrecy capacity is given by
\begin{align} \label{C_AN}
C_{\mathrm{AN}} &= {C_B} - {C_{E - \mathrm{AN}}}  \\ \notag
&\ge \log _2\left(\frac{1 + \delta \mu }{1 + \frac{\delta \mu \beta }{(1 - \delta )\mu \eta(1-\beta)  + \varepsilon }} \right)=C_{\mathrm{AN}}^{\mathrm{LB}},
\end{align}
where $C_{\mathrm{AN}}^{\mathrm{LB}}$ represents the lower bound of $C_{\mathrm{AN}}$.

To guarantee the transmission, i.e.,  $C_{\mathrm{AN}}^{\mathrm{LB}}\geq R_s$, where $R_s$ is a predefined secrecy transmission rate, we have
\begin{align} \label{beta_AN}
\beta\leq\frac{(1-\delta)\mu\eta+\varepsilon}{(1-\delta)\mu\eta+\frac{\delta\mu2^{R_s}}{1+\delta\mu-2^{R_s}}}.
\end{align}

Generally, the existing physical layer security techniques can be classified into two categories in terms of the effects on the eavesdropper. One is to reduce the information signal leakage, and the other is to enlarge interference dynamics through AN to disturb the received signal of eavesdropper. The latter approach unavoidably consume additional energy (e.g. for AN generation), and sacrifice the communication performance at legitimate user. To tackle this issue, we consider the DM scheme without AN in this part.

The DM without AN scheme (i.e., RFDA-DM scheme) is a special case of the scheme mentioned in II-B, thus means that the transmitting antenna will utilize all the transmit power to transmit confidential signal (i.e., $\delta  = 1$). As such, the lower bound of the secrecy capacity is given by
\begin{align}
C^{\mathrm{LB}}= \log _2\left(\frac{1 +  \mu }{1 + \frac{ \mu \beta }{ \varepsilon }} \right).
\end{align}

Likewise, \eqref{beta_AN} can be rewritten as
\begin{align} \label{beta_without_AN}
\beta\leq\frac{(1+\mu-2^{R_s})\varepsilon}{\mu2^{R_s}}.
\end{align}

The analysis for the upper bound of $\beta$ in \eqref{beta_AN} and \eqref{beta_without_AN} enables us to determine the minimum values of $M$ and $K$ in \eqref{M_eq} and \eqref{K_eq}, respectively.

\section{Numerical Results}
In this section, we use numerical simulation experiments to evaluate the performance of our proposed system. Without other statements, some main parameters used in this section are set as follows. The carrier frequency ${f_0}$ is set to 1 GHz (i.e., ${f_0} = 1$ GHz), the frequency increment is set to 1 MHz (i.e., $\Delta f = 1$ MHz), the element spacing is half of the wavelength (i.e., $d=c/2f_0$), the location of desired user is set at $(100\mathrm{m}, 45^ \circ)$, $\varepsilon  = 1$, and the noise variances of Bob and Eve are set to 0 dBm (i.e., $\sigma _B^2=\sigma _E^2= 0$ dBm). The offsets of angle and range offset are set as $5^\circ$ and $8\mathrm{m}$, respectively (i.e., $\Delta \theta  = {5^ \circ }, \Delta r = 8{\rm{m}}$). For the performance evaluation, we assume that the Eve's location is $(108\mathrm{m}, 40^ \circ)$, which is unknown to the transmitter. The power allocation factor is set as $\delta=0.6$, thus means that $60\%$ of the transmit power is allocated to the confidential signal and the rest is allocated to AN.

\begin{table*}[htbp]\scriptsize
  \centering
  \caption{Frequency increments for each elements of antenna array in Fig.~\ref{fig3}}
   \resizebox{18cm}{!}{\begin{tabular}{|c|c|c|c|c|c|c|c|c|c|c|c|c|c|c|c|c|c|c|}
    \hline
    \multicolumn{3}{|c|}{m} &
      1 &
      2 &
      3 &
      4 &
      5 &
      6 &
      7 &
      8 &
      9 &
      10 &
      11 &
      12 &
      13 &
      14 &
      15 &
      16
      \\
    \hline
    K=10405 &
      \multicolumn{2}{c|}{\multirow{3}{*}{\makecell[c]{Frequency Increment:\\${k_m}\Delta f$(MHz)}}} &
      -15.2 &
      7.4 &
      -36 &
      32.5 &
      18.5 &
      33.8 &
      -29.4 &
      4.5 &
      13.7 &
      17.4 &
      -56.2 &
      -17.8 &
      9.73 &
      17.8 &
      -22.9 &
      22.2
      \\
\cline{1-1}\cline{4-19}    K=12905 &
      \multicolumn{2}{c|}{} &
      -16.9 &
      8.2 &
      -40.1 &
      36.2 &
      20.6 &
      37.7 &
      -32.8 &
      5 &
      15.3 &
      19.4 &
      -62.6 &
      -19.8 &
      10.8 &
      19.8 &
      -25.5 &
      24.7
      \\
\cline{1-1}\cline{4-19}    K=15405 &
      \multicolumn{2}{c|}{} &
      -18.5 &
      8.96 &
      -43.8 &
      39.5 &
      22.5 &
      41.2 &
      -35.8 &
      5.47 &
      16.7 &
      21.2 &
      -68.4 &
      -21.6 &
      11.8 &
      21.7 &
      -27.9 &
      27
      \\
    \hline
    \end{tabular}}
  \label{tab:addlabel}
\end{table*}

\begin{figure} [t!]
    \begin{center}
    \includegraphics[width=3.5in, height=2.9in]{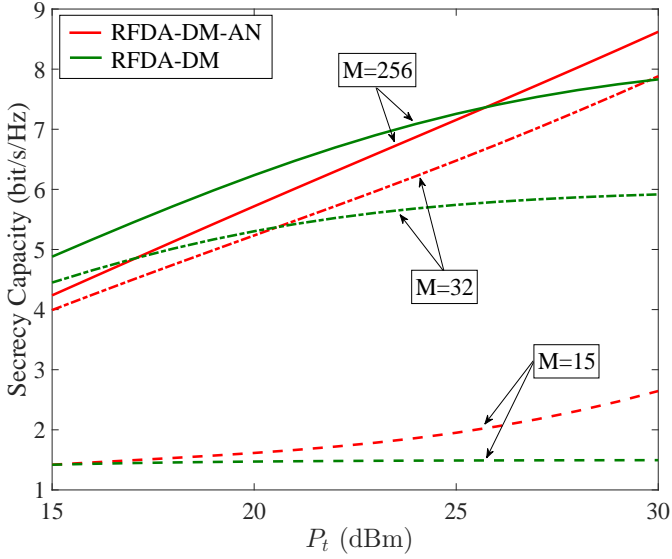}
    \caption{Secrecy capacity versus the transmit power $P_t$ under different values of $M$.}\label{fig2}
    \end{center}
\end{figure}

Fig.~\ref{fig2} plots the security capacity versus the transmit power $P_t$ under various values of $M$, where the minimum number of transmit antennas obtained from \eqref{M_eq}, $M_{\min}=15.7$ with $\beta=0.4$. In the Fig.~\ref{fig2}, we can observe that secrecy capacity of the RFDA-DM-AN scheme increases rapidly as $P_t$ increases, compared with RFDA-DM scheme, thus means that the secure transmission from Alice to Bob becomes easier when Alice allocate more transmit power to AN to enhance physical layer security. When $M=15<M_{\min}$, RFDA-DM-AN scheme completely outperforms RFDA-DM scheme, which confirms the correctness of our Theorem~\ref{theorem1}. In addition, as $M$ becomes larger (e.g., $M = 32,~256$), we observe that RFDA-DM scheme have a better performance than RFDA-DM-AN scheme when $P_t$ is smaller than a specified value, which is due to the fact that Alice can construct a narrow beam towards Bob when $M$ is large enough to avoid information leakage to Eve without the aid of AN when the transmit power is limited.

\begin{figure} [t!]
    \begin{center}
    \includegraphics[width=3.5in, height=2.9in]{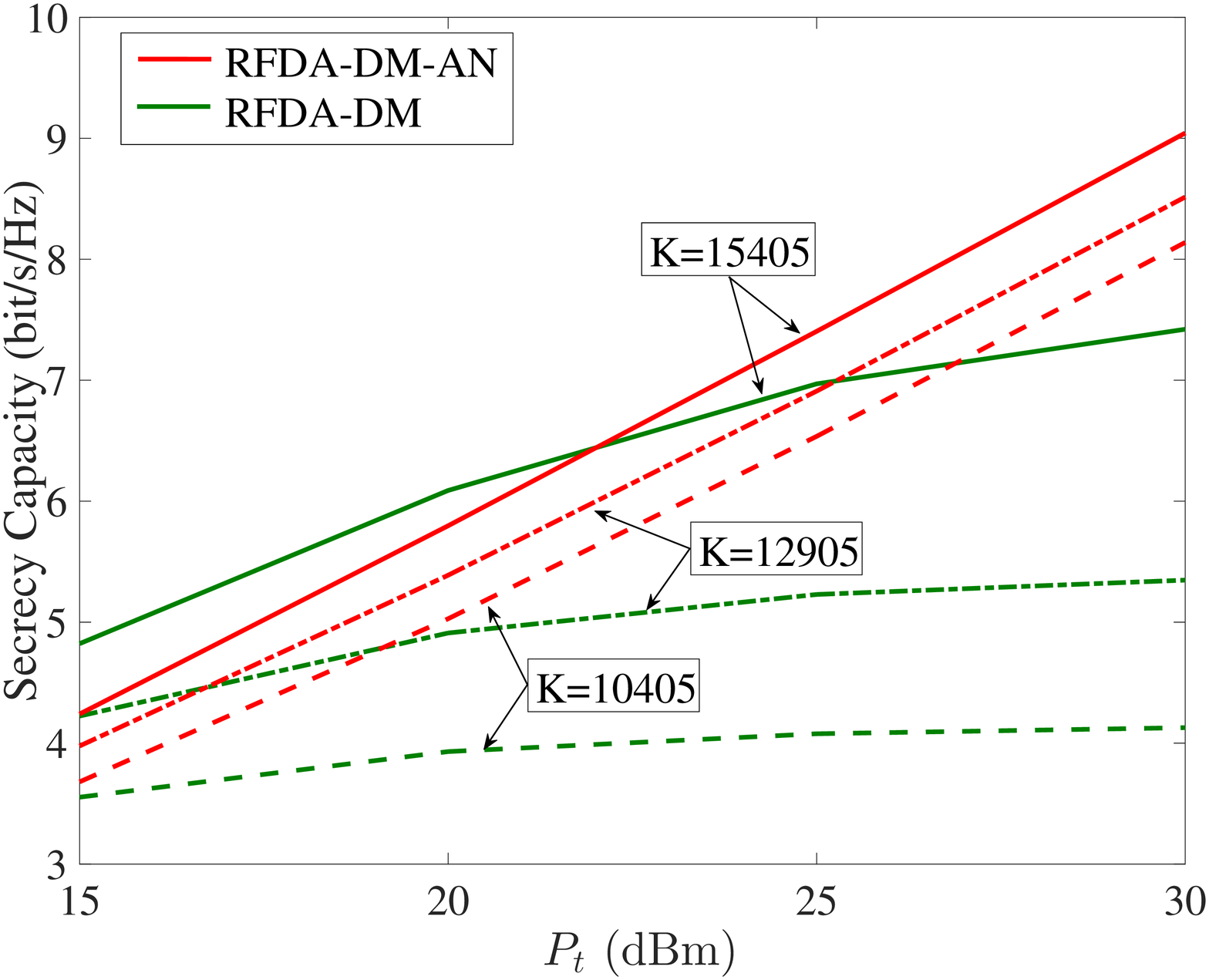}
    \caption{Secrecy capacity versus the transmit power $P_t$ under different values of $K$, where $M=16$.}\label{fig3}
    \end{center}
\end{figure}

Fig.~\ref{fig3} depicts the security capacity versus the transmit power $P_t$ under different values of $K$ for both RFDA-DM-AN and RFDA-DM schemes. We recall that $K$ is defined as $K = {\bf{k}}^T{{\bf{k}}}$, of which the value depends on the bandwidth allocated for frequency mapping.
Tab.~\ref{tab:addlabel} shows the frequency increments for each elements of the antenna array, and we can find the bandwidths for $K=10405, 12905, 15405$ are approximately equal to $90$~MHz, $100$~MHz, and $110$~MHz, respectively. In this figure, we observe that for both schemes security capacity monotonically increases as $K$ increases (equal increasing bandwidth). This is because the increasing bandwidth will enlarge the set for randomly allocating frequencies to improve the resolution in range dimension.

\begin{figure} [t!]
    \begin{center}
    \includegraphics[width=3.5in, height=2.9in]{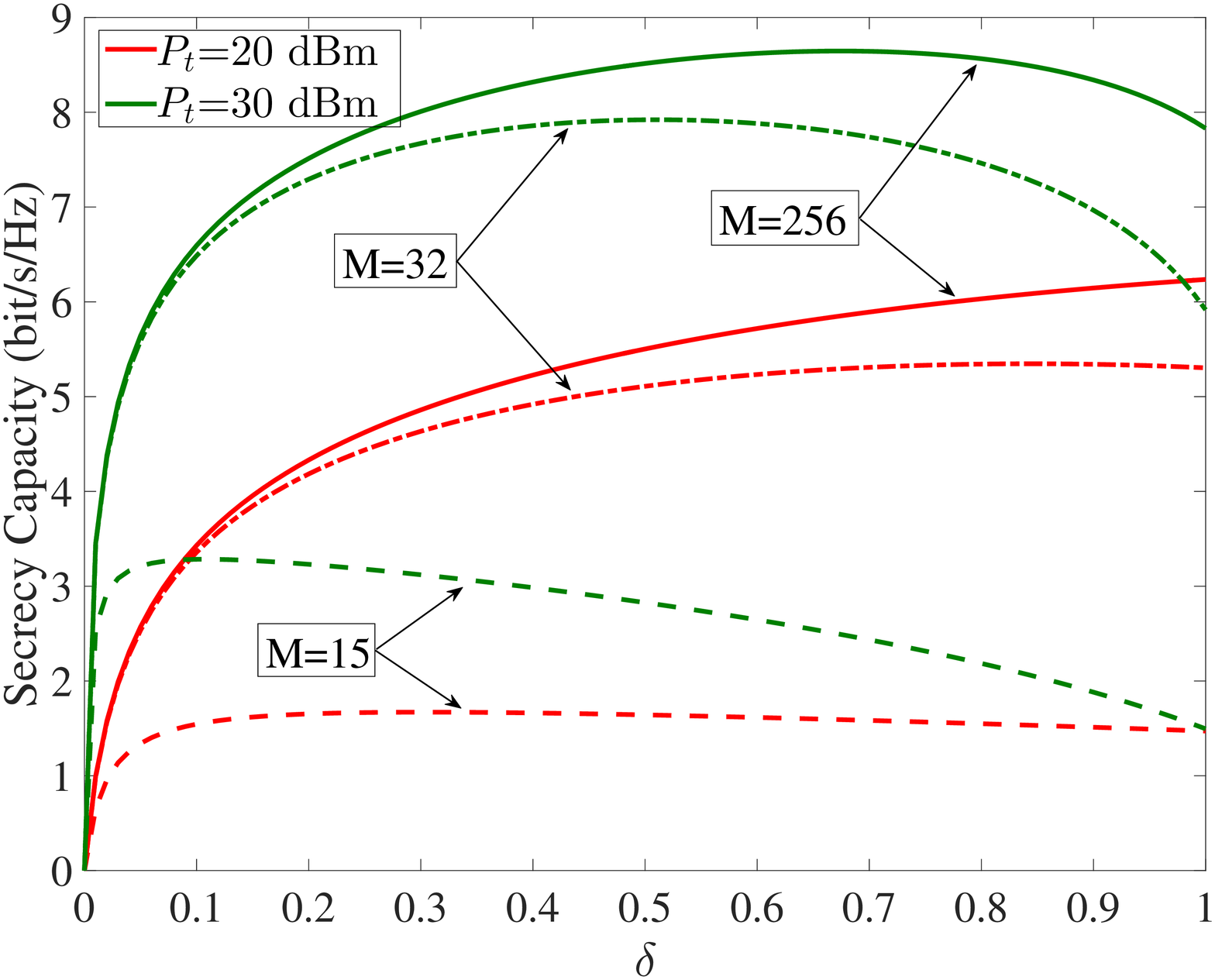}
    \caption{Secrecy capacity of the RFDA-DM-AN scheme versus the power allocation factor $\delta$ under different values of $M$.}\label{fig4}
    \end{center}
\end{figure}

Fig.~\ref{fig4} shows the effect of power allocation factor $\delta$ on the secrecy performance of the RFDA-DM-AN scheme. In this figure, we first observe that secrecy capacity first increases and then decreases as $\delta$ increases when the transmit power $P_t$ is sufficiently large (i.e., $P_t=30$~dBm), which indicates that there is an optimal value of $\delta$ that maximizes secrecy capacity. The optimal value of $\delta$ will increase when the transmitter are equipped with larger number of antennas, thus means that a greater proportion of the transmit power is allocated to the confidential signal in this case, which is due to the fact that the large number of antennas rather than AN guarantees the secure transmission.


\begin{figure} [t!]
    \begin{center}
    \includegraphics[width=3.5in, height=2.9in]{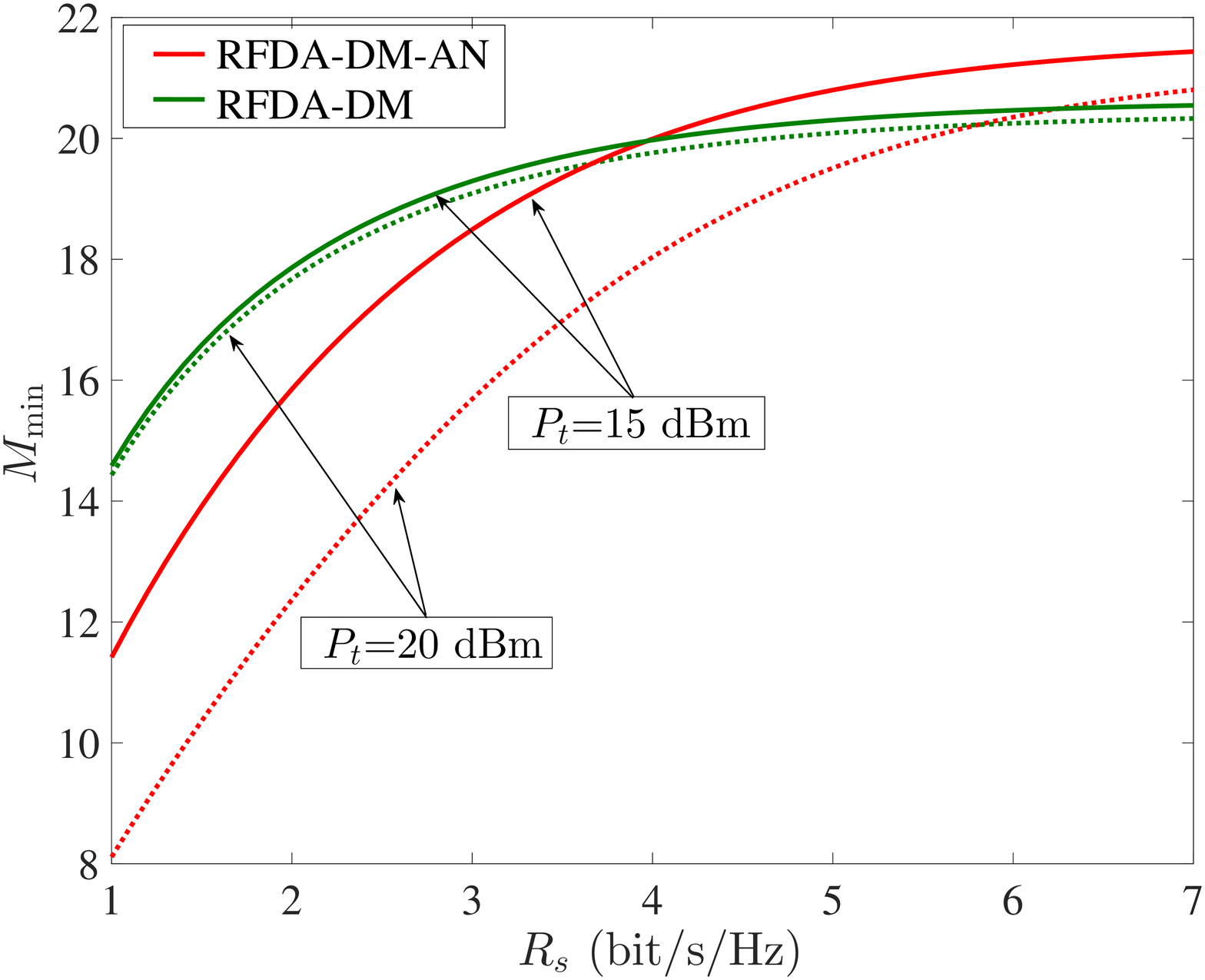}
    \caption{
    The minimum value of $M$ (i.e., $M_{\min}$) versus the predefined secrecy transmission rate $R_s$ under different values of $P_t$.}\label{fig5}
    \end{center} \vspace{-3mm}
\end{figure}

In Fig.~\ref{fig5}, we plot the minimum value of $M$ (i.e., $M_{\min}$) versus $R_s$ under different values of the transmit power $P_t$. From the figure, it can be observed that for both schemes $M_{\min}$ monotonically increases as $R_s$ increases. It implies that the transmitter requires a larger value of $M_{\min}$ to compensate the performance loss caused by the decrease of $P_t$.
We also note that $P_t$ has greater impact on $M_{\min}$ of RFDA-DM scheme than that of RFDA-DM-AN scheme for a given $R_s$.

\section{Conclusion}
This work examined secure communication with the aid of DM and RFDA, in which Alice transmits confidential signal to the desired user Bob, while Eve that may exist around Bob tries to wiretap this confidential signal. We theoretically analyzes the RFDA-DM and RFDA-DM-AN schemes with a given secrecy region. The beampattern of the RFDA around Bob can be translated into a ellipse function, based on which the closed forms of the minimum value of $M$ and $K$ to guarantee the secure communication and the secrecy capacities in RRFDA-DM and RFDA-DM-AN schemes are obtained. Numerical results have confirmed that the AN has a positive performance impact when $P_t$ is sufficiently large. Otherwise, without AN is a better choice. Moreover, we found that increasing the number of antennas could compensate the performance loss caused by a small $P_t$ or the absence of AN.

\bibliographystyle{IEEEtran}
\bibliography{IEEEfull,DM}

\end{document}